\documentclass[12pt]{article}
\pdfoutput=1
\usepackage[utf8]{inputenc}
\usepackage[T1]{fontenc}
\usepackage{float}
\usepackage{authblk}
\usepackage[normalem]{ulem}
\usepackage{amsfonts}       
\usepackage{amsmath}
\usepackage{amsthm}
\usepackage{amsbsy}
\usepackage{amssymb}
\usepackage{nicefrac}   
\usepackage{centernot}

\usepackage{adjustbox}
\usepackage{graphicx}
\usepackage{booktabs}
\usepackage{mathtools}
\usepackage{fullpage}
\usepackage[mathlines]{lineno}
\usepackage{subcaption}
\usepackage[style=numeric-comp,sorting=none,maxnames=99]{biblatex} 

\makeatletter
\renewcommand*\env@matrix[1][\arraystretch]{%
    \edef\arraystretch{#1}%
    \hskip -\arraycolsep
    \let\@ifnextchar\new@ifnextchar
    \array{*\c@MaxMatrixCols c}}
\makeatother

\usepackage{mathtools}

\usepackage{multirow}
\usepackage[table,xcdraw]{xcolor}

\usepackage[colorlinks = true,
            linkcolor = teal,
            urlcolor  = teal,
            citecolor = teal]{hyperref}

\title{Sparse coupling and Markov blankets \\ A comment on "How particular is the physics of the Free Energy Principle?" by Aguilera, Millidge, Tschantz and Buckley}

\author[1-4]{Conor Heins\thanks{\href{mailto:cheins@ab.mpg.de}{cheins@ab.mpg.de}}}
\author[5,6]{Lancelot Da Costa}

\affil[1]{Department of Collective Behaviour, Max Planck Institute \protect\\of Animal Behavior, 78464 Konstanz, Germany}
\affil[2]{Department of Biology, University of Konstanz, 78464 Konstanz, Germany}
\affil[3]{Centre for the Advanced Study of Collective Behaviour, \protect\\University of Konstanz, 78464 Konstanz, Germany}
\affil[4]{VERSES Research Labs, Los Angeles, California, USA}
\affil[5]{Department of Mathematics, \protect\\ Imperial College London, London SW7 2AZ, UK}
\affil[6]{Wellcome Centre for Human Neuroimaging, \protect\\University College London, London WC1N 3AR, UK}

\date{\vspace{-48pt}}
\begin{document}
\maketitle
\pagenumbering{arabic}


\begin{abstract}
    In this commentary, we respond to a technical analysis of the Free Energy Principle (hereafter: FEP) presented in "How particular is the physics of the Free Energy Principle" by Aguilera et al. In the target article, the authors analyzed certain sparsely coupled stochastic differential equations whose non-equilibrium steady-state densities are claimed---in previous FEP literature---to have a Markov blanket. The authors demonstrate that in general, Markov blankets are not guaranteed to follow from sparse coupling. The current commentary explains the relationship between sparse coupling and Markov blankets in the case of Gaussian steady-state densities. We precisely derive conditions under which causal coupling leads---or does not lead---to Markov blankets. Importantly, our derivations hold for both linear and non-linear stochastic differential equations. This result may shed light on the sorts of systems which we expect to have Markov blankets. Future work should focus on verifying whether these sorts of constraints are satisfied in realistic models of sparsely coupled systems.
\end{abstract}

\section{Overview}

The article "How particular is the physics of the Free Energy Principle?" \cite{aguilera2021particular} explores the different sorts of dynamical coupling structures in linear stochastic systems that do and do not entail conditional independence between states (i.e. Markov blankets \cite{da2021bayesian}) in the steady state distribution. The authors used these considerations to explore the limits and applicability of the Free Energy Principle (hereafter: FEP) in linear stochastic sytems. Previous FEP literature has intimated that sparse coupling structures (absent causal coupling between two subsets of states) lead to the emergence of Markov blankets \cite{friston2013life}. Further developments expanded on this statement to require additional sparsity constraints on the functional form of the flow \cite{friston2021stochastic,friston2019free}. 

The authors analyzed the `canonical loop' that is claimed to generically give rise to Markov blankets in \cite{friston2013life}. The authors show that in general Markov blankets are not guaranteed, and that `such a cyclic structure can generate couplings that propagate beyond causal interactions'. Although the exact relationship between sparse coupling and conditional independence was not rigorously defined in \cite{friston2013life}, subsequent work has constrained the implication of sparse coupling to conditional independence by invoking particular restrictions on the nature of the solenoidal coupling that connects different variables to one another \cite{friston2019free,friston2021some,friston2021stochastic}. The current commentary in many ways serves as a supplementary set of derivations that expands upon and explains mathematical relationships that have previously been shown.

In this commentary we precisely derive the conditions under which causal coupling leads---or does not lead---to conditional independence. Importantly, our derivations hold for both the linear case (the same class of systems investigated by the target article) as well as in the general, nonlinear case \cite{friston2021stochastic, friston2022free}. We show how conditional independence (Markov blankets) and causal coupling have a well-defined relationship that depends on the form of the solenoidal flow operator, which underlies the breaking of detailed balance \cite{tomita1974irreversible, eyink1996hydrodynamics, da2021bayesian}.

We show that the relationship between sparse coupling and conditional independence is particularly strengthened when one assumes the state-dependence of the solenoidal flow (a crucial feature underlying stochastic chaos and itinerancy in random dynamical systems \cite{friston2021stochastic, da2021bayesian}) has a `locality' constraint, whereby the coupling between any two pairs of states does not depend on other states besides the two that are interacting. This mathematical result may shed light on the sorts of systems we expect Markov blankets to emerge from sparse coupling. The `locality-of-coupling' constraint may particularly apply to the case of spatially-localized interactions (e.g. neighbouring particles or individuals interacting in a collective). The simple geometry of spatially-distributed systems characterized by short-range interactions may tend to enforce the locality conditions of solenoidal state-dependence and thus generally ensure that sparse coupling gives rise to Markov blankets. Future work, building along the lines of \cite{friston2021stochastic}, should focus on verifying whether these sorts of locality constraints on solenoidal dependence are satisfied in realistic models of locally-interacting systems.

\section{Sparse coupling in stochastic systems}

Most treatments of the FEP \cite{friston2013life, friston2019free, biehl2021technical, friston2022free} start by considering the evolution of a system defined over some state-space $x$, whose trajectories can be described using a Langevin-style (Itô) stochastic differential equation:
\begin{align}
    \dot{x}_t = f(x_t) + \omega
\end{align}

where $\dot{x}$ represents a partial temporal derivative and $\omega$ is additive Wiener noise.

We assume that the solution admits a (non-equilibrium) steady-state density $p$, i.e., a solution to the stationary Fokker-Planck equation \cite{pavliotis2014stochastic}
\begin{align}
    0 &= \nabla \cdot (\Gamma  \nabla p(x) - f(x)p(x)),
\end{align}
where $\Gamma = \mathbb E[\omega\omega^\top]/2$ is the diffusion tensor, expressing the covariance of random fluctuations. Thus, we can express the flow $f$ of the stochastic differential equation as a sum of dissipative (curl-free) and conservative (divergence-free) terms, using a Helmholtz decomposition \cite[Appendix B]{da2021bayesian}\footnote{See \cite[Appendix B]{da2021bayesian,eyink1996hydrodynamics} for a thermodynamic interpretation of these flows and \cite{barpUnifyingCanonicalDescription2021} for a geometric interpretation.}:
\begin{align}
    f(x) &= \underbrace{\Gamma(x) \nabla \log p(x) + \nabla \cdot \Gamma(x)}_{\text{dissipative}} - \underbrace{Q(x)\nabla \log p(x) - \nabla \cdot Q(x)}_{\text{conservative}}
\end{align}

Here and throughout, $\nabla \cdot Q$ denotes the divergence of the matrix field defined as $(\nabla \cdot Q)_{i}=\sum_{j} \frac{\partial}{\partial x_{j}} Q_{i j}$.\footnote{In other fields the same notation is used to denote the transpose of our definition \cite{deen2016introduction, noferesti2019numerical}, which amounts to a change of sign when the matrix field is antisymmetric.}
In the FEP literature, we often re-write the flow decomposition of $f(x)$ succinctly using the gradients of the negative log stationary density or surprisal  $\mathfrak{I}(x) = - \log p(x)$:
\begin{align}
    f(x) &= \boldsymbol \Omega(x) \nabla \mathfrak{I}(x) - \Lambda(x) \notag \\
    \textrm{where} \hspace{3.5mm} \boldsymbol \Omega(x) &\triangleq Q(x) - \Gamma(x), \quad \Lambda(x) \triangleq \nabla \cdot \boldsymbol \Omega(x).
\end{align}

To ease notation, we will omit writing the dependency on $x$ of $Q$ and $\boldsymbol \Omega$ in subsequent derivations, noting that this does not imply they are assumed to be constant, state-independent functions. However, we will assume constant, diagonal $\Gamma$ throughout. Imposing a constant, diagonal $\Gamma$ is the equivalent of forcing random fluctuations $\omega$ perturbing the system to be independently and identically distributed (i.i.d.).

Using this expression for the flow, we can then write the Jacobian of the system $\mathbf{J}$ where $\mathbf{J}_{uv} = \frac{\partial f_{u}}{\partial x_{v}}$ as follows:
\begin{align}
    \mathbf{J} &= \boldsymbol \Omega \mathbf{H} + \nabla \boldsymbol \Omega \nabla \mathfrak{I} - \nabla \Lambda. \label{eq:J_matrix_formula}
\end{align}

The Hessian $\mathbf{H}$ is a matrix whose entries encode the double partial derivatives (or `curvature') of the surprisal, with respect to the states: $\mathbf{H}_{uv} = \frac{\partial^2 \mathfrak{I}(x)}{\partial x_u \partial x_v}$. In the case of a Gaussian steady-state (i.e., a quadratic surprisal, where $\mathfrak{I}(x)$ is a second-order polynomial), then a zero-entry in $\mathbf{H}$ implies conditional independence and vice-versa \cite{friston2021stochastic,da2021bayesian}:
\begin{align}
    \mathbf{H}(x)_{u v} &=\frac{\partial^{2} \mathfrak{I}}{\partial x_{v} \partial x_{u}}=0  \notag \\
    & \Leftrightarrow \mathfrak{I}\left(x_{u} \mid b, x_{v}\right)=\mathfrak{I}\left(x_{u}|b\right) \notag \\
    & \Leftrightarrow \mathfrak{I}\left(x_{u}, x_{v}| b\right)=\mathfrak{I}\left(x_{u}| b\right)+\mathfrak{I}\left(x_{v} |b\right) \notag \\
    & \Leftrightarrow\left(x_{u} \perp x_{v}\right)|b: b=x_{\tilde{u}, \tilde{v}}
\end{align}

where the tilde notation $\tilde{u}$ denotes the complement of a set of states $u$. We can then use the matrix form of the Jacobian \eqref{eq:J_matrix_formula} to determine how the flow of one variable $x_u$ depends on the state of another $x_v$:

\begin{align}
    \mathbf{J}_{u v} &=\frac{\partial f_{u}}{\partial x_{v}}=\sum_{i} \boldsymbol \Omega_{u i} \mathbf{H}_{iv} + \sum_{i} \frac{\partial \boldsymbol \Omega_{u i}}{\partial x_{v}} \frac{\partial \mathfrak{I}}{\partial x_{i}} -\sum_{i} \frac{\partial^{2} \boldsymbol \Omega_{u i}}{\partial x_{i} \partial x_{v}}. \label{eq:single_J_entry}
\end{align}

When we assume a particular partition of the state-space $x$ into internal, active, sensory, and external states, respectively $x = (\mu, a, s, \eta)$, and further defining autonomous $\alpha = (\mu, a)$ and non-autonomous states $\beta =( s,  \eta) = \tilde{\alpha}$, we can then investigate the conditions under which sparse coupling is sufficient for conditional independence between variables, i.e., Markov blankets in the stationary density $p$.

One sort of coupling structure that can be related to conditional independence (under extra constraints on the flow operator), is the so-called `canonical loop' described in the target article \cite[Figure 4A]{aguilera2021particular}, which was introduced in earlier literature \cite{friston2013life, friston2021some, friston2021stochastic}. This structure assumes that sensory paths do not depend on internal paths, and active paths do not depend on external paths \cite{friston2022free}. This is best expressed in terms of the Jacobian:
\begin{align}
    \mathbf{J} &= \begin{bmatrix}
    \mathbf{J}_{\eta \eta} & \mathbf{J}_{\eta s} & \mathbf{J}_{\eta a} \\
    \mathbf{J}_{s \eta} & \mathbf{J}_{s s} & \mathbf{J}_{s a} & \\
    & \mathbf{J}_{a s} & \mathbf{J}_{a a} & \mathbf{J}_{a \mu} \\
    & \mathbf{J}_{\mu s} & \mathbf{J}_{\mu a} & \mathbf{J}_{\mu \mu}
    \end{bmatrix} \label{eq:canonical_J}
\end{align}

Given this sparse coupling structure and our expression for the Jacobian in terms of the Hessian in Equations \eqref{eq:J_matrix_formula} and \eqref{eq:single_J_entry}, one can then ask: what kinds of constraints must be in place on the flow operator $\boldsymbol \Omega$ in order for sparse coupling to imply conditional independence? By setting the left-hand side of Equation \eqref{eq:single_J_entry} to $0$, we can determine the conditions under which an absence of coupling from state $v$ to state $u$, implies a corresponding $0$ in the entry of the Hessian $\mathbf{H}_{uv}$.

\subsection*{The case of absent coupling}

To stay consistent with the target article's focus on linear Langevin equations, we first assume that $\boldsymbol \Omega$ and $\mathbf{H}$ are constant in the states. Thus Equation \eqref{eq:J_matrix_formula} reduces to:
\begin{align}
    \mathbf{J} = \boldsymbol \Omega \mathbf{H}
\end{align}

The usual correction terms $\nabla \boldsymbol \Omega \nabla \mathfrak{J}$ and $\nabla \Lambda$ disappear in the linear case because $\nabla_x \boldsymbol \Omega = 0$. We can expand the expressions for $\mathbf{J}_{\eta \mu}$ and $\mathbf{J}_{\mu \eta}$ to see what terms of $\boldsymbol \Omega$ and $\mathbf{H}$ they depend on:
\begin{equation}
\label{eq:expanded_etamu_J_entries}
\begin{split}
    \mathbf{J}_{\eta \mu} &= (Q_{\eta \eta} - \Gamma_{\eta})\mathbf{H}_{\eta \mu} + \sum_{i \in \tilde{\eta}} Q_{\eta i} \mathbf{H}_{i \mu}\\
    \mathbf{J}_{\mu \eta} &= (Q_{\mu \mu} - \Gamma_{\mu})\mathbf{H}_{\eta \mu}^\top + \sum_{i \in \tilde{\mu}} Q_{\mu i} \mathbf{H}_{i \eta} 
\end{split}
\end{equation}

Absent coupling in one direction or another enable us to derive specific expressions for the Hessian
\begin{align}
    \mathbf{J}_{\eta \mu}=0 &\iff \mathbf{H}_{\eta \mu}= (\Gamma_\eta-Q_{\eta \eta})^{-1}\sum_{i \in \tilde{\eta}} Q_{\eta i}\mathbf{H}_{i \mu}\\
    \mathbf{J}_{\mu\eta}=0 &\iff -\mathbf{H}_{\eta \mu}(Q_{\mu \mu} + \Gamma_{\mu}) - \sum_{i \in \tilde{\mu}}  \mathbf{H}_{\eta i }Q_{ i\mu} =0 \notag\\
    &\iff \mathbf{H}_{\eta \mu}= - (Q_{\mu \mu} + \Gamma_{\mu})^{-1}\sum_{i \in \tilde{\mu}}  \mathbf{H}_{\eta i }Q_{ i\mu}
\end{align}

Thus, we can completely characterise conditional independence under the absence of couplings in both directions:
\begin{equation}
\label{eq: cond indep, unidirectional coupling}
\begin{split}
        \text{Assuming } \mathbf{J}_{\eta \mu}=0: \mathbf{H}_{\eta \mu}=0 \iff \sum_{i \in \tilde{\eta}} Q_{\eta i}\mathbf{H}_{i \mu}=0\\
    \text{Assuming } \mathbf{J}_{\mu\eta}=0: \mathbf{H}_{\eta \mu}=0 \iff \sum_{i \in \tilde{\mu}}  \mathbf{H}_{\eta i }Q_{ i\mu}=0
\end{split}
\end{equation}

These expressions simplify  when we assume the so-called `normal' or `canonical' form of the flow \cite{friston2019free,friston2021some,friston2021stochastic,friston2022free}, in which the Jacobian is sparse because solenoidal couplings between autonomous ($\alpha = (\mu, a)$) and non-autonomous ($\beta = (\eta, s)$) states are assumed to be absent:

\begin{align}
    \begin{bmatrix}
    \mathbf{J}_{\eta \eta} & \mathbf{J}_{\eta s} & \mathbf{J}_{\eta a} \\
    \mathbf{J}_{s \eta} & \mathbf{J}_{s s} & \mathbf{J}_{s a} & \\
    & \mathbf{J}_{a s} & \mathbf{J}_{a a} & \mathbf{J}_{a \mu} \\
    & \mathbf{J}_{\mu s} & \mathbf{J}_{\mu a} & \mathbf{J}_{\mu \mu}
    \end{bmatrix} &= \begin{bmatrix}
    Q_{\eta \eta}-\Gamma_{\eta} & Q_{\eta s} & & \\
    -Q_{\eta s}^\top & Q_{s s}-\Gamma_{s} & & \\
    & & Q_{a a}-\Gamma_{a} & Q_{a \mu} \\
    & & -Q_{a \mu}^\top & Q_{\mu \mu}-\Gamma_{\mu}
    \end{bmatrix}\begin{bmatrix}
    \mathbf{H}_{\eta \eta} & \mathbf{H}_{\eta s} & \mathbf{H}_{\eta a} & \mathbf{H}_{\eta \mu} \\
    \mathbf{H}_{\eta s}^\top & \mathbf{H}_{s s} & \mathbf{H}_{s a} & \mathbf{H}_{s \mu} \\
    \mathbf{H}_{\eta a}^\top & \mathbf{H}_{s a}^\top & \mathbf{H}_{a a} & \mathbf{H}_{a \mu} \\
    \mathbf{H}_{\eta \mu}^\top & \mathbf{H}_{s \mu}^\top & \mathbf{H}_{a \mu}^\top & \mathbf{H}_{\mu \mu}
    \end{bmatrix} \label{eq:normal_form_linear}
\end{align}

Note that the `block-solenoidal' terms $Q_{\eta \eta}$ and $Q_{\mu \mu}$ are $0$ if each state-subset is univariate (e.g. $\mu, \eta$ are both one-dimensional).

Assuming the canonical form of the flow \eqref{eq:normal_form_linear}, certain terms in \eqref{eq: cond indep, unidirectional coupling} vanish and the conditional independence condition is simplified
\begin{align}
\label{eq:H_etamu_constraint_normal_form}
\text{Assuming } \eqref{eq:normal_form_linear} : \mathbf{H}_{\eta \mu}=0 \iff Q_{\eta s}\mathbf{H}_{s \mu}=0 \iff \mathbf{H}_{\eta a}Q_{a \mu}=0.
\end{align}
For example, if we assume absent off-diagonal solenoidal flow, then Markov blankets are guaranteed from sparse coupling.

An open question, which the authors of the target article touched on in their analytic expressions for the drift matrices (or Jacobians) of linear systems, is whether linear systems that satisfy conditional independence under sparse coupling are `rare' or not. The classification of a particular linear system as `rare' depends on an underlying assumption about the distribution of (sparse) drift matrices from which we expect such diffusions to be generated. In particular, if we assume that both statistical and solenoidal coupling in real systems is local, then both $Q$ and $\mathbf{H}$ will become sparser as dimension increases, and the entries of the inner products in \eqref{eq:H_etamu_constraint_normal_form} will vanish with high probability. This can be easily shown by observing that the probability of sparse vectors being orthogonal (their inner-product vanishing) asymptotically approaches $1.0$ as dimension increases. Whether it is fair to assume locality (and thus sparsity) in the $Q$ and $\mathbf{H}$ operators, depends on prior assumptions about the nature of solenoidal coupling and covariance relationships that characterize real systems with Gaussian steady state \cite{dempster1972covariance, friedman2008sparse}. For example, we may expect solenoidal operators $Q$ to be sparse for complex systems defined by localized (and thus sparse) interactions across a network of spatially-segregated components.

In the more general non-linear case where $\nabla_x \boldsymbol \Omega(x) \neq 0$, additional terms (corresponding to the summands of the correction term $\Lambda(x)$) appear in \eqref{eq:expanded_etamu_J_entries} and propagate into \eqref{eq: cond indep, unidirectional coupling} and \eqref{eq:H_etamu_constraint_normal_form}.

\subsection*{The sparse coupling conjecture}
We now examine the conditions under which one can impute conditional independence from not only absent coupling $\mathbf{J}_{u v} = \mathbf{J}_{vu} = 0$, but also from \textit{unidirectional or non-reciprocal} coupling, i.e., $\mathbf{J}_{u v} = 0, \mathbf{J}_{vu} \neq 0$. The so-called \textit{sparse coupling conjecture} \cite{friston2021stochastic, friston2022free} conjectures the following implication in a generic class of systems:
\begin{align}
    \mathbf{J}_{uv}\mathbf{J}_{vu} &= 0 \implies \mathbf{H}_{uv} = \mathbf{H}_{uv} = 0.
\end{align}

Expanding the case of $\mathbf{J}_{\eta a} \neq 0, \mathbf{J}_{a \eta} = 0$ from \eqref{eq:normal_form_linear}, we can see why this implies conditional independence between $\eta$ and $a$. If assume there is a Markov blanket between internal and external states (i.e. the constraints in \eqref{eq:H_etamu_constraint_normal_form} hold) then we can simplify the expressions for the coupling between $\eta$ and $a$:
\begin{align}
    \text{Assuming } \mathbf{H}_{\eta \mu} = 0: 
    \mathbf{J}_{\eta a} &= (Q_{\eta \eta} - \Gamma_\eta)\mathbf{H}_{\eta a} + \sum_{i \in \tilde{\eta}} Q_{\eta i}\mathbf{H}_{i a} \notag \\
    &= \left(Q_{\eta \eta} - \Gamma_\eta\right)\mathbf{H}_{\eta a} + Q_{\eta s}\mathbf{H}_{s a} \notag \\
    \mathbf{J}_{a \eta} &= (Q_{a a} - \Gamma_a)\mathbf{H}_{a \eta}^\top + \sum_{i \in \tilde{a}} Q_{a i} \mathbf{H}_{i \eta}= 0 \notag \\
    &=\left(Q_{a a} - \Gamma_{a}\right)\mathbf{H}_{a \eta}^\top
    \label{eq:J_eta_a_reduced}
\end{align}

The only way for $\mathbf{J}_{a \eta}$ to vanish while $\mathbf{H}_{a \eta}^\top = \mathbf{H}_{\eta a}$ is non-zero, is for $Q_{a a}\mathbf{H}_{a \eta}^{\top} = \Gamma_{a}\mathbf{H}_{a \eta}^{\top}$, which cannot hold because these terms are by definition orthogonal to another and can only be equal if they both vanish, which is impossible due to the the positive-definiteness of $\Gamma_{a}$ and $\mathbf{H}$. Therefore $\mathbf{H}_{a \eta}^\top = \mathbf{H}_{\eta a} = 0$ and the sparse coupling conjecture is verified for the normal form laid out in Equation \eqref{eq:normal_form_linear}. Similar reasoning can be used to expand the entries for $\mathbf{J}_{s \mu}$ and $\mathbf{J}_{\mu s}$ to re-write the normal form with a fully-sparse Hessian and Jacobian:
\begin{align}
    \begin{bmatrix}
    \mathbf{J}_{\eta \eta} & \mathbf{J}_{\eta s} & \mathbf{J}_{\eta a} \\
    \mathbf{J}_{s \eta} & \mathbf{J}_{s s} & \mathbf{J}_{s a} & \\
    & \mathbf{J}_{a s} & \mathbf{J}_{a a} & \mathbf{J}_{a \mu} \\
    & \mathbf{J}_{\mu s} & \mathbf{J}_{\mu a} & \mathbf{J}_{\mu \mu}
    \end{bmatrix} &= \begin{bmatrix}
    Q_{\eta \eta}-\Gamma_{\eta} & Q_{\eta s} & & \\
    -Q_{\eta s}^\top & Q_{s s}-\Gamma_{s} & & \\
    & & Q_{a a}-\Gamma_{a} & Q_{a \mu} \\
    & & -Q_{a \mu}^\top & Q_{\mu \mu}-\Gamma_{\mu}
    \end{bmatrix} \begin{bmatrix}
    \mathbf{H}_{\eta \eta} & \mathbf{H}_{\eta s} &  \\
    \mathbf{H}_{\eta s}^\top & \mathbf{H}_{s s} & \mathbf{H}_{s a} &  \\
    & \mathbf{H}_{s a}^\top & \mathbf{H}_{a a} & \mathbf{H}_{a \mu} \\
    & & \mathbf{H}_{a \mu}^\top & \mathbf{H}_{\mu \mu}
    \end{bmatrix}.
\end{align}

Now, for generality, we refine these conditions for the fully nonlinear case, assuming for simplicity that the only component of the flow operator that is state-dependent is $Q$ ($\Gamma$ is state-independent and diagonal, as before):
\begin{align}
    \mathbf{J}(x) &= (Q - \Gamma) \mathbf{H}+\nabla Q \cdot \nabla \mathfrak{J}-\nabla \Lambda\\
    \begin{bmatrix}
    \mathbf{J}_{\eta \eta} & \mathbf{J}_{\eta s} & \mathbf{J}_{\eta a} \\
    \mathbf{J}_{s \eta} & \mathbf{J}_{s s} & \mathbf{J}_{s a} & \\
    & \mathbf{J}_{a s} & \mathbf{J}_{a a} & \mathbf{J}_{a \mu} \\
    & \mathbf{J}_{\mu s} & \mathbf{J}_{\mu a} & \mathbf{J}_{\mu \mu}
    \end{bmatrix} &= \begin{bmatrix}
    Q_{\eta \eta}-\Gamma_{\eta} & Q_{\eta s} & & \\
    -Q_{\eta s}^\top & Q_{s s}-\Gamma_{s} & & \\
    & & Q_{a a}-\Gamma_{a} & Q_{a \mu} \\
    & & -Q_{a \mu}^\top & Q_{\mu \mu}-\Gamma_{\mu}
    \end{bmatrix}\begin{bmatrix}
    \mathbf{H}_{\eta \eta} & \mathbf{H}_{\eta s} & \mathbf{H}_{\eta a} & \mathbf{H}_{\eta \mu} \\
    \mathbf{H}_{\eta s}^\top & \mathbf{H}_{s s} & \mathbf{H}_{s a} & \mathbf{H}_{s \mu} \\
    \mathbf{H}_{\eta a}^\top & \mathbf{H}_{s a}^\top & \mathbf{H}_{a a} & \mathbf{H}_{a \mu} \\
    \mathbf{H}_{\eta \mu}^\top & \mathbf{H}_{s \mu}^\top & \mathbf{H}_{a \mu}^\top & \mathbf{H}_{\mu \mu}
    \end{bmatrix} \notag \\
    &+ \begin{bmatrix}
    \nabla_{\eta} Q \\
    \nabla_{s} Q \\
    \nabla_{a} Q \\
    \nabla_{\mu} Q
    \end{bmatrix} \cdot\begin{bmatrix}
    \nabla_{\eta} \mathfrak{I}(x) \\
    \nabla_{s} \mathfrak{I}(x) \\
    \nabla_{a} \mathfrak{I}(x) \\
    \nabla_{\mu} \mathfrak{I}(x)
    \end{bmatrix}
    - \begin{bmatrix}
    (\nabla \Lambda)_{\eta \eta} & (\nabla \Lambda)_{\eta s} & (\nabla \Lambda)_{\eta a} & (\nabla \Lambda)_{\eta \mu} \\
    -(\nabla \Lambda)_{\eta s}^\top & (\nabla \Lambda)_{s s} & (\nabla \Lambda)_{s a} & (\nabla \Lambda)_{s \mu} \\
    -(\nabla \Lambda)_{\eta a}^\top & -(\nabla \Lambda)_{s a}^\top & (\nabla \Lambda)_{a a} & (\nabla \Lambda)_{a \mu} \\
    -(\nabla \Lambda)_{\eta \mu}^\top & -(\nabla \Lambda)_{s \mu}^\top & -(\nabla \Lambda)_{a \mu}^\top & (\nabla \Lambda)_{\mu \mu}
    \end{bmatrix} \label{eq:full_jacobian_nonlinear}
\end{align}

Where the gradient of the correction term $\nabla \Lambda$ can be re-written as:
\begin{equation}
    (\nabla \Lambda)_{uv} = \nabla (\nabla \cdot Q) = \sum_{i} \frac{\partial^{2} Q_{u i}}{\partial x_{i} \partial x_{v}}.
\end{equation}

We can now derive explicit conditions on the nature of the solenoidal flow's state dependence that guarantee, barring edge cases, an implication of conditional independence from sparse coupling. We once again assume a Markov blanket between $\eta$ and $\mu$, i.e. the nonlinear equivalent of \eqref{eq:H_etamu_constraint_normal_form} is satisfied.

Taking coupling between internal and sensory states as an example, the sparse coupling in Equation \eqref{eq:full_jacobian_nonlinear} assumes $\mathbf{J}_{s \mu} = 0$. We can expand its expression in the Jacobian, now including the gradients and curvatures of the solenoidal flow:
\begin{align}
    \mathbf{J}_{s \mu} = \left(Q_{s s} - \Gamma_{ss}\right) \mathbf{H}_{s \mu} + \nabla_{\mu} Q_{s \eta}\cdot \nabla_{\eta}\mathfrak{I}(x) + \nabla_{\mu} Q_{s s}\cdot \nabla_{s}\mathfrak{I}(x) - \nabla_{s \mu}^2 Q_{s s} - \nabla_{\eta \mu}^2 Q_{\eta s}. \label{eq:expanded_J_example}
\end{align}

The final two terms $\nabla_{s \mu}^2 Q_{s s}$ and $\nabla_{\eta \mu}^2 Q_{s \eta}$ vanish under the assumption that the solenoidal flow depends at most linearly on `non-local' terms.\footnote{By `non-local' dependence we mean if the solenoidal coupling $Q_{uv}$ between two states $u$ and $v$ depends on some third state $y$, i.e. $\partial_y Q_{uv} \neq 0$.} If we restrict this flow further, such that $Q_{s \eta}$ \textit{only} depends on $s$ and $\eta$, then all the partial derivatives of the solenoidal flow with respect to other states (here, $\mu$) vanish, meaning the expression for the Jacobian reduces to:
\begin{align}
    \mathbf{J}_{s \mu} = \left(Q_{s s} - \Gamma_{ss}\right) \mathbf{H}_{s \mu}. \label{eq:reduced_J_example}
\end{align}

Therefore $\mathbf{J}_{s \mu}$ will be 0 if and only if $s$ and $\mu$ are conditionally independent, i.e., $\mathbf{H}_{s\mu} = 0$.

If we however allow the solenoidal flow to have `non-local' state-dependence, then the presence of the remaining partial derivatives in \eqref{eq:expanded_J_example} could render the normal form a counter-example to the sparse coupling conjecture, unless the extra terms all cancel. This means that even in the case of conditional independence between $s$ and $\mu$, the coupling between $s$ and $\mu$ could be non-zero, if the solenoidal flow depends on more states than only those that it `connects.' This has interesting implications for the constraints we might expect in natural systems, where the interactions mediated by solenoidal flow should be `locally-dependent' (only depend on the states that are interacting) if Markov blankets are to be implied by sparse coupling. This may be a natural tendency in the case of spatially-extensive systems (e.g. cells, neurons, collectives of individuals) where localized interactions may be expected to only depend on states that are participating in the interaction.

Assuming edge-cases are not satisfied, we can now re-write Equation \eqref{eq:full_jacobian_nonlinear} with the flow respecting the assumption of local state-dependence, and the Hessian expressing more stringent conditional independence relationships:

\begin{align}
    \begin{bmatrix}
    \mathbf{J}_{\eta \eta} & \mathbf{J}_{\eta s} & \mathbf{J}_{\eta a} \\
    \mathbf{J}_{s \eta} & \mathbf{J}_{s s} & \mathbf{J}_{s a} & \\
    & \mathbf{J}_{a s} & \mathbf{J}_{a a} & \mathbf{J}_{a \mu} \\
    & \mathbf{J}_{\mu s} & \mathbf{J}_{\mu a} & \mathbf{J}_{\mu \mu}
    \end{bmatrix} &= \boldsymbol \Omega \begin{bmatrix}
    \mathbf{H}_{\eta \eta} & \mathbf{H}_{\eta s} &  \\
    \mathbf{H}_{\eta s}^\top & \mathbf{H}_{s s} & \mathbf{H}_{s a} &  \\
    & \mathbf{H}_{s a}^\top & \mathbf{H}_{a a} & \mathbf{H}_{a \mu} \\
    & & \mathbf{H}_{a \mu}^\top & \mathbf{H}_{\mu \mu}
    \end{bmatrix}
    + \begin{bmatrix}
    \nabla \boldsymbol \Omega_{\eta \eta} & \nabla \boldsymbol \Omega_{\eta s} & & \\
    -\nabla \boldsymbol \Omega_{\eta s}^\top & \nabla \boldsymbol \Omega_{s s} & & \\
    & & \nabla \boldsymbol \Omega_{a a} & \nabla \boldsymbol \Omega_{a \mu} \\
    & & -\nabla \boldsymbol \Omega_{a \mu}^\top & \nabla \boldsymbol \Omega_{\mu \mu}
    \end{bmatrix} \cdot \nabla \mathfrak{I}
    \notag \\
    &- \begin{bmatrix}
    (\nabla \Lambda)_{\eta \eta} & (\nabla \Lambda)_{\eta s} & & \\
    -(\nabla \Lambda)_{\eta s}^\top & (\nabla \Lambda)_{s s} & & \\
    & & (\nabla \Lambda)_{a a} & (\nabla \Lambda)_{a \mu} \\
    & & -(\nabla \Lambda)_{a \mu}^\top & (\nabla \Lambda)_{\mu \mu}
    \end{bmatrix}
     \label{eq:final_jacobian}
\end{align}

where the state-dependent flow operator $\boldsymbol \Omega = Q - \Gamma$, as before, lacks coupling between autonomous and non-autonomous states as in Equation \eqref{eq:full_jacobian_nonlinear}.

\section{Concluding remarks}

We hope the derivations provided here in addition to those in the target article contribute to an improved understanding of the relationship between dynamical and statistical dependence in Langevin equations with a steady-state.

Our derivations suggest that the lack of coupling between two states only implies conditional independence in certain cases, as specified in Equation \eqref{eq:H_etamu_constraint_normal_form}. In that sense, we find general agreement with the target article that in general, causal coupling does not have a straightforward relationship to conditional independence or the presence of Markov blankets. However, we note that in the limit of increasing dimensionality and sparse $Q$ and $\mathbf{H}$, then Markov blankets are likely to be implied by sparse coupling, specifically because the inner products in \eqref{eq: cond indep, unidirectional coupling} are expected to vanish, by virtue of being equivalent to the dot products (or sums thereof) of sparse vectors. We also show (in agreement with statements made in \cite{friston2021stochastic,friston2022free}) that even \emph{unidirectional coupling} can imply conditional independence, but further constraints on the flow operator and the nature of its state-dependence are required in order for this to hold true. Interestingly, the functional form of the state-dependence of the solenoidal flow, that is required in order to sustain conditional independence from sparse coupling, may agree with the sorts of `locality constraints' we would expect in natural systems, where any nonlinear coupling between interacting states should only depend on the states participating in the interaction. This result may provide a formal basis for the `local' nature of interactions in spatially-extended, self-organizing systems, whose state-space can often be decomposed into causally-coupled sub-systems.



We conclude by celebrating the words of the authors of the target article, on how these sorts of investigations offer an opportunity to study in detail "the connection between non-trivial dynamics and the statistical properties of coupled systems"~\cite{aguilera2021particular}. 

\section*{Additional information}\label{sec:additional_info}

\subsection*{Acknowledgements}\label{sec:ack}
The authors would like to thank Miguel Aguilera for pointing out a technical error in an earlier preprinted version of this commentary. We would also like to thank Dalton Sakthivadivel, Maxwell Ramstead, Karl Friston, and Magnus Koudahl for helpful discussions and feedback that improved the content of this commentary.  

\subsection*{Author contributions}

CH: Conceptualization, Formal analysis, Writing - Original Draft. LD: Conceptualization, Writing - Review \& Editing.

\subsection*{Funding statement}\label{sec:fund}
CH is supported by the U.S. Office of Naval Research (N00014-19-1-2556), and acknowledges the support of a grant from the John Templeton Foundation (61780). The opinions expressed in this publication are those of the author(s) and do not necessarily reflect the views of the John Templeton Foundation.

LD is supported by the Fonds National de la Recherche, Luxembourg (Project code: 13568875). This publication is based on work partially supported by the EPSRC Centre for Doctoral Training in Mathematics of Random Systems: Analysis, Modelling and Simulation (EP/S023925/1).

\printbibliography[title={References}]

\end{document}